\documentclass[prd,twocolumn,tightenlines,superscriptaddress,nofootinbib]{revtex4}
\usepackage{amssymb,latexsym}
\usepackage{amsmath,amsbsy,bbm}
\usepackage{epsfig,bm}
\usepackage{graphicx,comment}
\usepackage{color}
\usepackage[normalem]{ulem}
\begin{document}

\title{Quantum phase transitions of 2-d dimerized spin-1/2 Heisenberg models with spatial anisotropy}

\author{M.-T. Kao}
\author{D.-J. Tan}
\author{F.-J. Jiang}
\email[]{fjjiang@ntnu.edu.tw}
\affiliation{Department of Physics, National Taiwan Normal University, 
88, Sec.4, Ting-Chou Rd., Taipei 116, Taiwan}

\vspace{-2cm}

\begin{abstract}

Motivated by the unexpected Monte Carlo results as well as 
the theoretical proposal of a large correction to scaling for
the critical theory of the 2-d staggered-dimer spin-1/2 Heisenberg model 
on the square lattice, we study the phase transitions induced by dimerization
of several dimerized quantum Heisenberg models with spatial anisotropy
using first principles Monte Carlo method. 
Remarkably, while our Monte Carlo data for all the models 
considered here, including the
herringbone- and ladder-dimer models on the square lattice,
are compatible with the recently proposed scenario of an enhanced correction to scaling, 
we find it is likely that the enhanced correction to scaling manifests itself as
amplification of the nonuniversal prefactors appeared in the scaling forms. 
In other words, our data are in consistence with the established 
numerical values for the critical exponents, including the confluent exponent,
in the $O(3)$ universality class. Convincing numerical evidence is provided to
support this proposed scenario. 

\end{abstract}


\maketitle

\section{Introduction}
While being well-studied and understood thoroughly, the dimerized quantum Heisenberg models 
with spatial anisotropy have triggered theoretical interests again recently 
\cite{Matsumoto01,Tro02,Wang05,Yas05,Hastings06,Praz06,Yao07,Hog071,Hog072,Pardini08,Jiang09.1,Sandvik11.10}. For example, 
the 3-d spatially anisotropic quantum Heisenberg model with a ladder dimerization pattern 
is used to demonstrate a universal behavior, which is argued
to be relevant for understanding the experimental results of the material
TlCuCl$_3$ \cite{Oitmaa11.1}. Further, the 2-d dimerized spin-1/2 Heisenberg model 
with a spatially staggered anisotropy is of particularly interesting because
this model seems to establish an unconventional phase transition 
\cite{Wenzel08}. Specifically, although it is
believed that the phase transition induced by dimerization for this model 
should be governed by the $O(3)$ universality class
theoretically \cite{Chakr88,Haldane88,Chubu94,Sachdev99,Vojta03},
a recent large scale Monte Carlo calculation obtains $\nu = 0.689(5)$ and
$\beta/\nu = 0.545(5)$, which are in contradiction to the established $O(3)$
results $\nu = 0.7112(5)$ and $\beta/\nu = 0.519(1)$ in the literature \cite{Cam02}. 
Here $\nu$ and $\beta$ are the critical exponents corresponding to
the correlation length and the magnetization, respectively.
In order to clarify this issue further, several efforts have been devoted to 
study
the phase transition of this model induced by dimerization. For instance, an 
unconventional finite-size scaling is proposed in \cite{Jiang11.8}. Further, 
in \cite{Fritz11} it is 
argued that, due to a cubic term, there is a large correction to scaling for this phase 
transition which results in the unexpected $\nu = 0.689(5)$ and
$\beta/\nu = 0.545(5)$ obtained in \cite{Wenzel08}. 
Later, a Monte Carlo study indeed provides strong evidence to 
support this scenario of an enhanced correction to scaling \cite{Jiang11.8}.
In addition to the staggered-dimer spin-1/2 Heisenberg model on the square lattice, 
in \cite{Fritz11} it concludes as well that 
a similar model on the honeycomb lattice, which is depicted in the bottom panel of 
figure 1\footnote{We will call this model the staggered-dimer model as well 
unless confusion arises.}, as well as the herringbone-dimer model
on the square lattice (middle panel of figure 1) also belong to the same category 
of models receiving an enhanced correction. This general picture 
regarding the correction to scaling for 2-d dimerized quantum
Heisenberg models is indeed supported by several related Monte Carlo
studies \cite{Matsumoto02,Wang05,Albu08,Wenzel09,Jiang09.2}.

While all the available Monte Carlo results provide 
convincing evidence for the proposal of an enhanced correction to scaling, 
the good scaling property of the observable $\rho_{s2}2L$ for
the staggered-dimer model on the square lattice is 
the most noticeable observation \cite{Jiang11.8}, 
where again $\rho_{s2}$ and $L$ are the spin stiffness in the
2-direction and the spatial box size employed in the simulations, respectively. Inspired by 
this observation, one naturally would like to examine whether for the staggered-dimer model 
on the honeycomb lattice and the herringbone-dimer model on the square lattice, 
a similar good scaling behavior will be observed when considering the same observable 
$\rho_{s2}2L$ for these two dimerized models. Although the phase transition 
induced by dimerization of the staggered-dimer 
model on the honeycomb lattice has been studied before, a detailed
comparison between the scaling behavior of $\rho_{s1}2L$ and $\rho_{s2}2L$ as well as
the relevant investigation of the exponent $\beta/\nu$ are not available 
yet \cite{Wenzel09.1}.
In addition, to examine how the enhanced correction 
to scaling, due to a cubic irrelevant term as suggested in \cite{Fritz11}, 
affects the determination of the exponents $\nu$ and
$\beta/\nu$ for the staggered- and herringbone-dimer models is an interesting topic 
to explore as well \cite{Wenzel09.2}. 
Indeed, whether the cubic term will influence the numerical value of the confluent exponent $\omega$
has not explored in \cite{Fritz11}. Hence in this study, we have investigated the phase transitions of 
the herringbone- and ladder-dimer spin-1/2 Heisenberg models on the square 
lattice, as well as the quantum staggered-dimer model on the honeycomb lattice. In particular,
the largest lattice sizes reached here are as twice large as those of the relevant early 
studies in some cases. The results for the ladder-dimer model are included here for completeness 
and comparison purpose, since the enhanced correction to scaling should be absent for this model. 
Remarkably, as we will demonstrate later, indeed $\rho_{s2}2L$ of the staggered-dimer
model on the honeycomb lattice shows a good scaling behavior. 
Consequently, we are able to obtain a value for $\nu$, in agreement with the established 
$\nu=0.7112(5)$ in the $O(3)$ universality class, by employing the leading scaling
ansatz in our finite-size scaling analysis for $\rho_{s2}2L$.
Interestingly, while our Monte Carlo data for all the models studied here, including the
herringbone- and ladder-dimer models on the square lattice, are
compatible with the recently proposed scenario of an enhanced correction to scaling
for the phase transitions considered here, we find that the enhanced correction to scaling 
manifests itself as amplification of the nonuniversal prefactors appeared in the scaling forms. 
In other words, our data are in consistence with the established 
results of $\nu =0.7112(5)$, $\beta/\nu = 0.519(1)$, and 
$\omega \sim 0.78$ in the $O(3)$ universality class. We provide
numerical evidence to support this proposed scenario of amplification of the 
nonuniversal prefactors appeared in the scaling forms, 
by demonstrating that the values of $\omega$ for the herringbone- and
staggered-dimer models are compatible with $\omega \sim 0.78$, through a calculation of 
relevant observables at the corresponding critical points.

This paper is organized as follows. First, after an introduction, 
the spatially anisotropic quantum
Heisenberg models and the relevant observables studied in this work are briefly described, 
after which we 
present our numerical results. In particular, the results obtained from the 
finite-size scaling analysis are discussed in detail.
A final section then concludes our study.

\section{Microscopic Model and Corresponding Observables}
The Heisenberg
models considered in this study are defined by the Hamilton operator
\begin{eqnarray}
\label{hamilton}
H = \sum_{\langle xy \rangle}J\,\vec S_x \cdot \vec S_{y}
+\sum_{\langle x'y' \rangle}J'\,\vec S_{x'} \cdot \vec S_{y'},
\end{eqnarray}
where $J$ and $J'$ are antiferromagnetic exchange couplings connecting
nearest neighbor spins $\langle  xy \rangle$
and $\langle x'y' \rangle$, respectively. Figure 1 illustrates 
the models which are described by Eq.~(\ref{hamilton}) and are
investigated in great detail here.
To study the critical behavior of these models near 
the transition driven by the anisotropy, in particular, to determine 
the critical points as well as the critical exponent $\nu$, 
the spin stiffnesses in the $1$- and $2$-directions which are defined by\vskip-0.5cm
\begin{eqnarray}
\rho_{si} = \frac{1}{\beta L^2}\langle W^2_{i}\rangle,
\end{eqnarray}
are measured in our simulations.
Here $\beta$ is the inverse temperature and $L$ again 
refers to the spatial box size. Further $\langle W^2_{i} \rangle$ 
with $i \in \{1,2\}$ is
the winding number squared in the $i$ direction.
In addition, the second Binder ratio $Q_2$, which is defined by
\begin{equation}
Q_2 = \frac{\langle (m_s^z)^2\rangle^2}{\langle (m_s^z)^4\rangle}
,\end{equation}
is also measured in our simulations as well. Here $m_s^z$ is 
the $z$ component of the staggered magnetization 
$\vec{m}_s = \frac{1}{L^2}\sum_{x}(-1)^{x_1 + x_2}\vec{S}_x$.
By carefully investigating the spatial volume 
and the $J'/J$ dependence of
$\rho_{s i}L$ as well as $Q_2$, one can determine the critical points and the critical 
exponent $\nu$ with high precision.
Finally the exponent $\beta/\nu$ is determined 
by studying the scaling behavior of the observables $\langle |m_s^z| \rangle$
and $\langle (m_s^z)^2\rangle$, which are measured in this study as well, at 
the corresponding critical points.

\section{Determination of the Critical Points and the Critical Exponent $\nu$} 

To study the quantum phase transitions of our central interest,
we have carried out large scale Monte Carlo simulations using a loop algorithm 
\cite{Evertz93.1,Evertz93.2,Wiese94,Beard96,Troyer08}. 
Further, to calculate the relevant critical exponent $\nu$ and to determine the
location of the critical points in the parameter space $J'/J$ for the models
described by figure 1, we have employed 
the technique of finite-size scaling for certain observables. For example,
if a transition is second order, then at low-temperature\footnote{Specifically, one
expects that the temperature should be lower than the energy gap $\Delta \sim 1/L$ 
for the systems considered in this study.} 
and near the transition the observable 
$\rho_{si} 2L$ for $i\in \{1,2\}$ and $Q_2$ should be described well by the following finite-size scaling ansatz 
\cite{Fisher72,Brezin82,Barber83,Brezin85,Fisher89}
\begin{eqnarray}
\label{FSS}
{\cal O}_{L}(t) &=& g_{{\cal O}}(tL^{1/\nu},\Delta^{-1}/\beta,r)+L^{-\omega}g_{{\cal O_{\omega}}}(tL^{1/\nu},\Delta^{-1}/\beta,r)\nonumber \\
              &=& g_{{\cal O}}(tL^{1/\nu},\Delta^{-1}/\beta,r) \times \nonumber \\
  &&\left( 1 + L^{-\omega} g'_{{\cal O_{\omega}}}(tL^{1/\nu},\Delta^{-1}/\beta,r)\right)
\end{eqnarray}
where ${\cal O}_{L}$ stands for $Q_2$ and $\rho_{si}L$ with $i\in\{1,2\}$, $L$ is the lattice size in the
1-direction, $t = (j_c-j)/j_c$ with $j = (J'/J)$,
$\nu$ is the critical exponent corresponding to the correlation length $\xi$,
$\omega$ is the confluent correction exponent, $\Delta$ is the energy gap which scales as 
$\Delta \sim 1/L^z$ with $z$ being
the dynamical critical exponent (which is $1$ for the phase 
transitions considered here), and $r$ is the ratio of the lattice size
in the 1- and 2-direction.
Further, $g_{{\cal O}}$, $g_{{\cal O_{\omega}}}$, and $g'_{{\cal O_{\omega}}}$ appearing above are
smooth functions of the variables $tL^{1/\nu}$, $\Delta^{-1}/\beta$, and $r$.
In practice one would carry out the
analysis close to the critical point so that 
$g'_{{\cal O_{\omega}}}$ in Eq.~(\ref{FSS}) can be approximated by a constant. Specifically,    
the following ansatz
\begin{equation}
\label{approx}
{\cal O}_{L}(t) = (1+bL^{-\omega})g_{{\cal O}}(tL^{1/\nu},L^{z}/\beta,r),
\end{equation}
where $b$ is some constant, is frequently used when applying the finite-size scaling technique. 
While Eq.~(\ref{approx}) is only valid for large box sizes and close to the critical point,
to present the main
results of this study we find it is sufficient to employ
Eq.~(\ref{approx}) for the data analysis. Notice that for square lattice or 
rectangular-shape lattice with a fixed $r$, 
one will intuitively neglect the effect of $r$ in Eq.~(\ref{approx}). 
Hence, we will apply Eq.~(\ref{approx}) with a constant $r$ to the relevant observables
for obtaining $(J'/J)_c$ and $\nu$. Notice from Eq.~(\ref{approx}), 
one concludes that the curves for ${\cal O}_{L}$ corresponding to different $L$, as functions of $J'/J$, should intersect 
at the critical point $(J'/J)_c$ for large $L$. 
Without loss of generality, we have 
fixed $J=1$ in our simulations and have varied $J'$. Additionally, the box size used in 
the simulations ranges from $L = 24$ to $L = 136$ (Strictly speaking, $L=\sqrt{N}$ for the staggered-dimer model
on the honeycomb lattice. Here $N$ is the number of spins used in the simulations).
Notice to eliminate the temperature dependence in Eq.~(\ref{approx}),
one naturally would use large enough inverse temperature $\beta$ in the simulations so that all the considered 
observables
take their zero-temperature values. On the other hand, since Eq.~(\ref{approx})
is valid for sufficiently low temperature, one can optimize the ratio of $\beta$ and $L$
in order to reach a lattice size as large as possible. As a result,
we use $\beta J = 2L$ for each $L$ in our simulations so that the temperature dependence
in Eq.~(\ref{approx}) drops out. We have generated some data using lower temperature
and these new data points lead to consistent results with those determined by employing
the data obtained with $\beta J = 2L$. First of all, let us focus on our results for the staggered-dimer
spin-1/2 Heisenberg model on the honeycomb lattice.

\begin{figure}
\label{fig1}
\begin{center}
\vbox{
\includegraphics[width=0.32\textwidth]{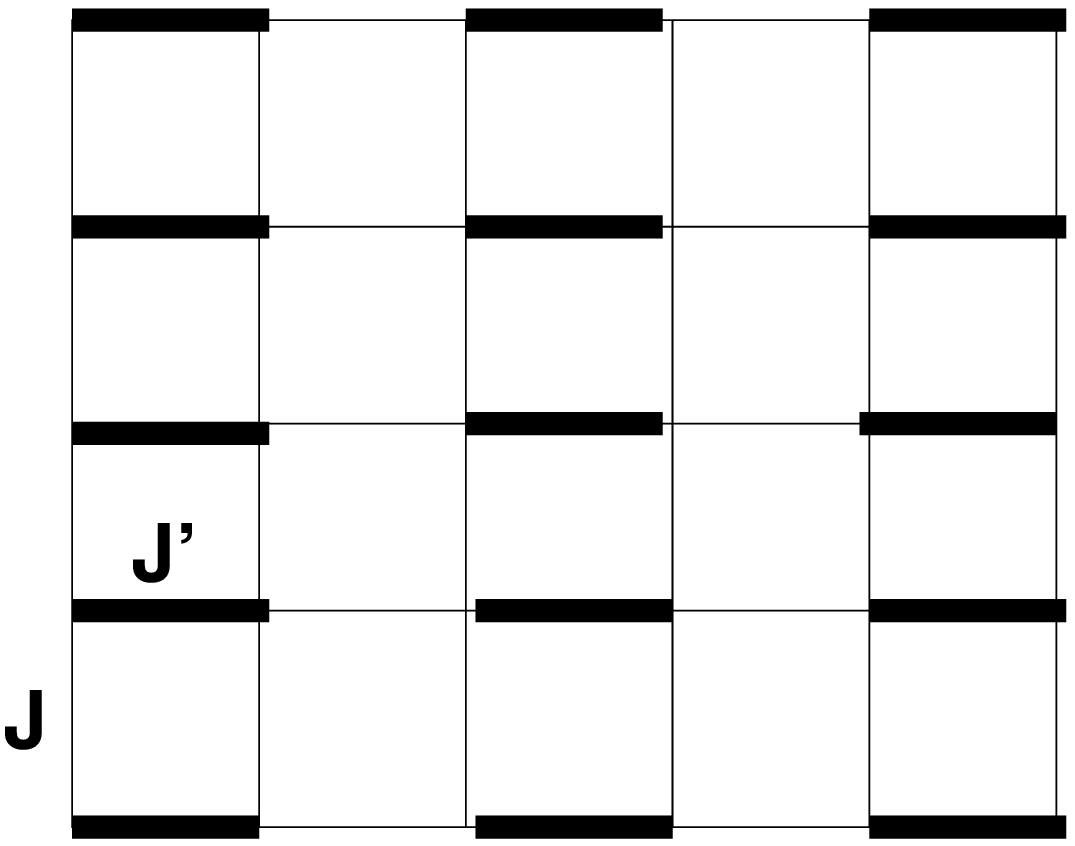}~~
\vskip0.5cm
\includegraphics[width=0.33\textwidth]{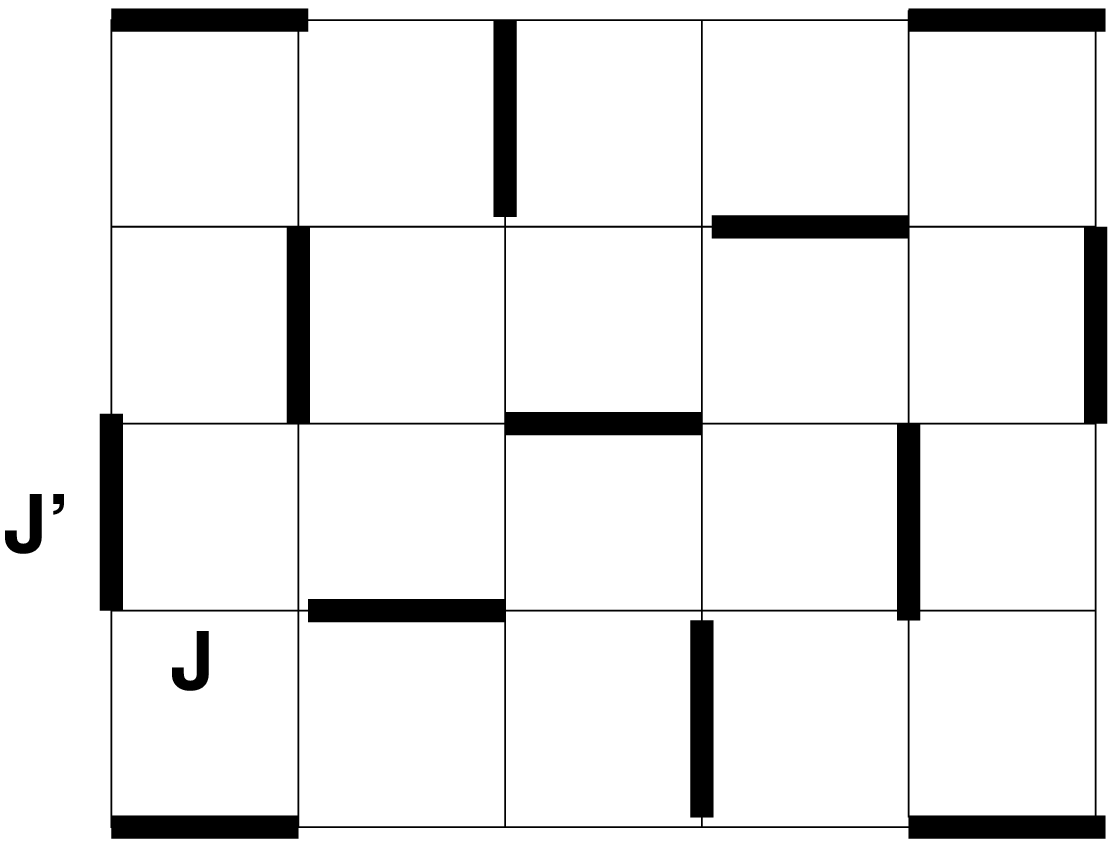}~~~
\vskip0.5cm
\includegraphics[width=0.32\textwidth]{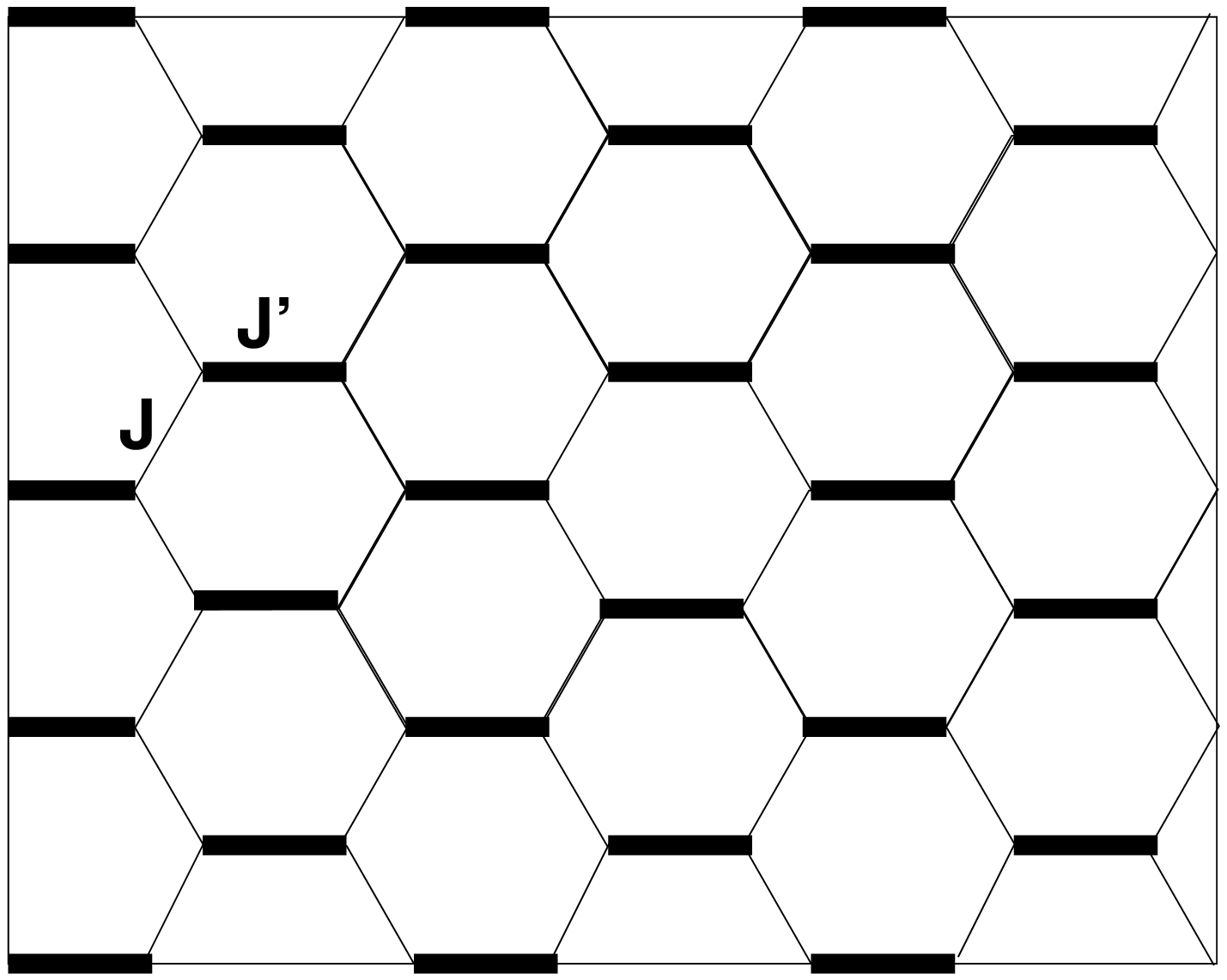}
}
\end{center}
\caption{The dimerized quantum Heisenberg models with spatial anisotropy 
considered in this study.
}
\end{figure}

\begin{figure}
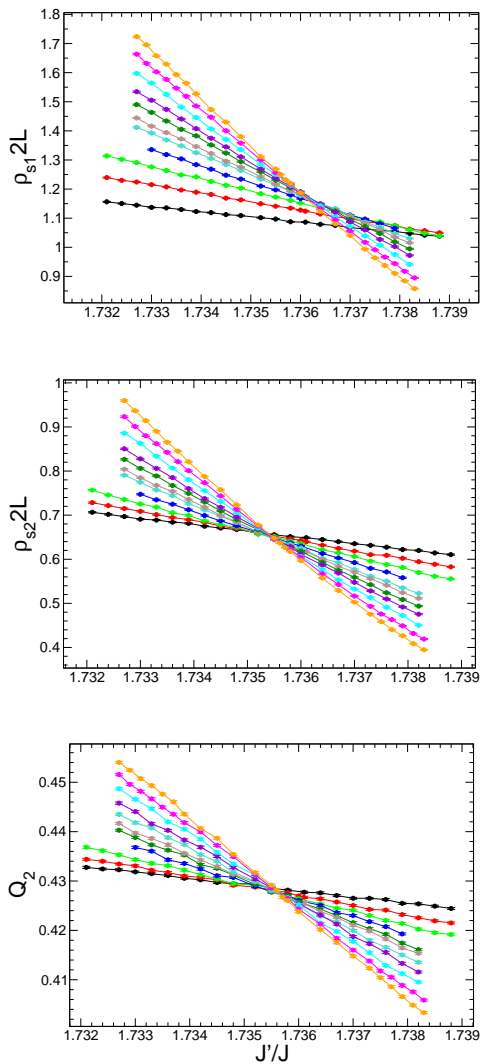

\label{fig2}
\begin{center}
\vbox{
\includegraphics[width=0.35\textwidth]{honeycomb_staggered_rhos12L_1.eps}
\vskip0.45cm
\includegraphics[width=0.35\textwidth]{honeycomb_staggered_rhos22L_1.eps}
\vskip0.47cm
\includegraphics[width=0.35\textwidth]{honeycomb_staggered_Q2_1.eps}
}
\end{center}
\caption{Monte Carlo data of $\rho_{s1}2L$, $\rho_{s2}2L$, and $Q_2$ with $24 \le L \le 96$ for 
the staggered-dimer spin-1/2 Heisenberg model on the honeycomb lattice. 
}
\end{figure}

\subsection{Results for the staggered-dimer spin-1/2 Heisenberg model on the honeycomb lattice}
Figure 2 shows the Monte Carlo data 
of $\rho_{s1} 2L$, $\rho_{s2} 2L$, and $Q_2$ with $24 \le L \le 96$ as functions of $J'/J$ for the staggered-dimer
spin-1/2 Heisenberg model on the honeycomb lattice. 
The figure clearly indicates that the phase 
transition is most likely second order since for all the observables $\rho_{s1} 2L$, $\rho_{s2} 2L$, and $Q_2$, 
the curves of different $L$ tend to intersect near a particular point in 
the parameter space $J'/J$. A surprising observation from figure 2 is that, while
$\rho_{s1}2L$ receives a sizable correction to its scaling (which has already been 
shown in \cite{Jiang09.2}), the observable $\rho_{s2}2L$ shows a good scaling behavior. 
Specifically, the correction to scaling
for $\rho_{s2}2L$ is negligible for $L \ge 32$.
These findings are similar to the scenario regarding the correction to scaling for 
the same observables, namely $\rho_{s1}2L$ and
$\rho_{s2}2L$, of the staggered-dimer model on the square lattice \cite{Jiang11.8}. 
Indeed, from $\rho_{s2}2L$ with $L \ge 32$, we are able to reach a value 
for $\nu$ compatible with
the expected $\nu=0.7112(5)$ using the leading finite-size scaling ansatz in Eq.~(\ref{approx})
(letting $b=0$ in Eq.~(\ref{approx})). For example, the $\nu$ obtained
from applying a second order Taylor expansion in $tL^{1/\nu}$ of Eq.~(\ref{approx}) with $b = 0$ to the 
observable $\rho_{s2}2L$ with $40 \le L \le 96$ 
is given by $\nu = 0.7167(40)$, which is in nice agreement with its theoretical expectation in the 
literature. To reach a value for $\nu$ consistent with $\nu = 0.7112(5)$ using the leading
finite-size scaling ansatz and the observable $\rho_{s1}2L$, one has to use data with 
fairly large $L$ as indicated in \cite{Jiang09.2}. Indeed a similar conclusion is reached here. 
Interestingly, with the observable $\rho_{s1}2L$, while we either arrive at values of $\nu$ 
statistically different from $\nu = 0.7112(5)$ or cannot reach good results when $b$ and $\omega$ in Eq.~(\ref{approx}) 
are included as fitting parameters, compatible results of $\nu$ with 
$\nu=0.7112(5)$ can be obtained from the fits with the assumption that $\omega \le 0.5$ is used as
a criterion for the fits. For instance, using $\rho_{s1}2L$ with $40 \le L \le 96$, the values of 
$\nu$ and $\omega$ determined from the fits 
by employing the criterion of $\omega \le 0.5$ are
given by $\nu=0.7054(45)$ and $\omega = 0.42(8)$, respectively. Notice $\omega \sim 0.42$ is smaller than the expected
$\omega \sim 0.78$ in the $O(3)$ universality class. One might conclude that,
our results are consistent with the scenario outlined in \cite{Fritz11} 
that the correction to scaling for this model is enhanced due to a cubic 
irrelevant term, and this term has impact on the numerical value of $\omega$.
Finally, from $Q_2$ with $32 \le L \le 96$, a fit using 
Taylor expansion to second
order in $tL^{1/\nu}$ as well as letting $b=0$ in Eq.~(\ref{approx}) leads to
$\nu=0.7102(56)$ and $(J'/J)_c = 1.73560(4)$, both of which agree 
quantitatively with the known results in the literature. 
Interestingly, the obtained 
coefficients for $(tL^{1/\nu})^2$ in the fits associated with $Q_2$ are very small. 
Hence, we can even reach 
a value for $\nu$ in agreement with $\nu=0.7112(5)$ using a first order Taylor expansion
in $tL^{1/\nu}$ of Eq.~(\ref{approx}) to fit the data points of $Q_2$ (the last 2 rows in table 1).
The values of $\nu$ and $(J'/J)_c$ obtained from the fits mentioned above are listed in table 1. 
Notice that the uncertainties of $(J'/J)_c$ and $\nu$ shown in table 1, as well as in tables 
2, 3, 4, and 5 in the following sections, are determined by a conservative estimate based on 
the standard deviations obtained from the bootstrap resampling method employed for the fits. 

\begin{table}
\label{tab1}
\begin{center}
\begin{tabular}{ccccc}
\hline
{\text{observable}}& $\,L\, $  & $\nu$ & $(J'/J)_c$ & $\chi^2/{\text{DOF}}$\\
\hline
\hline
$\rho_{s1}2L$  & $40\, \le \,L\, \le\, 96$  & 0.7054(45) & 1.7355(2) & 1.2\\
\hline
$\rho_{s1}2L$  & $48\, \le \,L\, \le\, 96$  & 0.7096(55) & 1.7355(3) & 1.2\\
\hline
$\rho_{s2}2L$  & $32\, \le \,L\, \le\, 96$  & 0.7156(40) & 1.73545(3) & 1.9\\
\hline
$\rho_{s2}2L$  & $40\, \le \,L\, \le\, 96$  & 0.7167(40) & 1.73548(3) & 1.4\\
\hline
$\rho_{s2}2L$  & $48\, \le \,L\, \le\, 96$  & 0.7055(42)$^{\star}$ & 1.73550(3)$^{\star}$ & 0.9\\
\hline
$\rho_{s2}2L$  & $56\, \le \,L\, \le\, 96$  & 0.7082(45)$^{\star}$ & 1.73551(3)$^{\star}$ & 0.9\\
\hline
$Q_2\,$&  $32\, \le \,L\, \le\, 96$  & 0.7102(56) & 1.73560(4) & 1.4\\
\hline
$Q_2\,$&  $40\, \le \,L\, \le\, 96$  & 0.712(6) & 1.73570(5) & 1.3\\
\hline
$Q_2\,$&  $32\, \le \,L\, \le\, 96$  & 0.7107(56)$^{\diamond}$ & 1.73564(3)$^{\diamond}$  & 1.4\\
\hline
$Q_2\,$&  $40\, \le \,L\, \le\, 96$  & 0.712(6)$^{\diamond}$  & 1.73566(4)$^{\diamond}$  & 1.4\\
\hline
\hline
\end{tabular}
\end{center}
\caption{The numerical values of $\nu$ and $(J'/J)_c$ calculated from 
$\rho_{s1}2L$, $\rho_{s2}2L$, 
and $Q_2$ for the staggered-dimer model on the honeycomb lattice. 
All results are obtained by using a second order Taylor expansion in $tL^{1/\nu}$ of 
Eq.~(\ref{approx}) except those with a star (diamond) which are determined by a third 
order (first order) Taylor expansion.
The confluent correction is included in the fit explicitly only for $\rho_{s1}2L$ and 
is assumed to satisfy the condition $\omega \le 0.5$.
}
\end{table}

\begin{table}
\label{tab2}
\begin{center}
\begin{tabular}{ccccc}
\hline
{\text{observable}}& $\,L\, $  & $\nu$ & $(J'/J)_c$ & $\chi^2/{\text{DOF}}$\\
\hline
\hline
$\rho_{s}L$  & $24\, \le \,L\, \le\, 136$  & 0.705(2) & 2.49804(8) & 1.4\\
\hline
$\rho_{s}L$  & $24\, \le \,L\, \le\, 96$  & 0.706(3) & 2.4980(2) & 1.4 \\
\hline
$\rho_{s}L$  & $24\, \le \,L\, \le\, 72$  & 0.702(5) & 2.4980(5) & 1.1 \\
\hline
$\rho_{s}L$  & $32\, \le \,L\, \le\, 136$  & 0.706(2) & 2.49805(10) & 1.4\\
\hline
$\rho_{s}L$  & $32\, \le \,L\, \le\, 96$  & 0.707(3) & 2.4980(3) & 1.4 \\
\hline
$\rho_{s}L$  & $24\, \le \,L\, \le\, 72$  & 0.700(4)$^{\star}$ & 2.49813(10)${^\star}$ & 1.1 \\
\hline
$\rho_{s}L$  & $32\, \le \,L\, \le\, 136$  & 0.706(2)$^{\star}$ & 2.49806(3)${^\star}$ & 1.5 \\
\hline
$\rho_{s}L$  & $32\, \le \,L\, \le\, 96$  & 0.707(3)$^{\star}$ & 2.49806(7)${^\star}$ & 1.4 \\
\hline
$Q_2$  & $24\, \le \,L\, \le\, 136$  & 0.714(4) & 2.49800(7) & 0.9\\
\hline
$Q_2$  & $32\, \le \,L\, \le\, 136$  & 0.715(6) & 2.4980(2) & 1.1\\
\hline
$Q_2$  & $24\, \le \,L\, \le\, 136$  & 0.710(4)$^{\star}$ & 2.49820(6)${^\star}$ & 1.1\\
\hline
$Q_2$  & $24\, \le \,L\, \le\, 96$  & 0.716(5)$^{\star}$ & 2.4983(1)${^\star}$ & 1.1\\
\hline
$\rho_{s}L$  & $24\, \le \,L\, \le\, 136$  & 0.701(2) & 2.49803(7) & 1.2 \\
\hline
$\rho_{s}L$  & $32\, \le \,L\, \le\, 136$  & 0.702(3) & 2.4980(1) & 1.2 \\
\hline
$Q_2$  & $24\, \le \,L\, \le\, 136$  & 0.707(4) & 2.49800(15) & 0.8\\
\hline
$Q_2$  & $32\, \le \,L\, \le\, 136$  & 0.7075(45) & 2.4980(2) & 0.9\\
\hline
$Q_2$  & $32\, \le \,L\, \le\, 136$  & 0.7056(40)$^{\star}$ & 2.49810(6)${^\star}$ & 0.9\\
\hline
$Q_2$  & $40\, \le \,L\, \le\, 136$  & 0.7065(40)$^{\star}$ & 2.49810(7)${^\star}$ & 0.8\\
\hline
\hline
\end{tabular}
\end{center}
\caption{The numerical values of $\nu$ and $(J'/J)_c$ calculated from 
$\rho_{s}L$ and $Q_2$ for the herringbone-dimer model on the square lattice. 
While the results presented in the first twelve rows are obtained by using a second 
order Taylor expansion in $tL^{1/\nu}$ of Eq.~(\ref{approx}), those listed in the last
six rows are determined with a third order Taylor expansion. 
Further, all results are calculated with the $\omega$ and $b$ in Eq.~(\ref{approx}) 
left as fitting parameters except those with a star which are determined through 
fits with a fixed $\omega = 0.78$. }
\end{table}

\begin{table}
\label{tab2.5}
\begin{center}
\begin{tabular}{ccccc}
\hline
{\text{observable}}& $\,L\, $  & $\nu$ & $\omega$ & $\chi^2/{\text{DOF}}$\\
\hline
\hline
$\rho_{s}L$  & $24\, \le \,L\, \le\, 136$  & 0.708(5) & 0.53(3) & 1.6\\
\hline
$\rho_{s}L$  & $32\, \le \,L\, \le\, 136$  & 0.710(5) & 0.57(5) & 1.7 \\
\hline
$\rho_{s}L$  & $40\, \le \,L\, \le\, 136$  & 0.711(6) & 0.63(8) & 1.7 \\
\hline
$\rho_{s}L$  & $24\, \le \,L\, \le\, 96$  & 0.706(8) & 0.48(5) & 1.6\\
\hline
$\rho_{s}L$  & $24\, \le \,L\, \le\, 136$  & 0.711(7)$^{\star}$ & {\text{N/A}} & 2.0 \\
\hline
$\rho_{s}L$  & $32\, \le \,L\, \le\, 136$  & 0.711(6)$^{\star}$ & {\text{N/A}} & 1.8 \\
\hline
$Q_2$  & $24\, \le \,L\, \le\, 136$  & 0.710(5) & 2.2(1) & 0.9 \\
\hline
$Q_2$  & $32\, \le \,L\, \le\, 136$  & 0.710(5) & 2.42(25) & 0.9 \\
\hline
$Q_2$  & $24\, \le \,L\, \le\, 96$  & 0.712(7) & 2.10(13) & 1.0\\
\hline
$Q_2$  & $24\, \le \,L\, \le\, 136$  & 0.714(9)$^{\star}$ & {\text{N/A}} & 1.8\\
\hline
$Q_2$  & $32\, \le \,L\, \le\, 136$  & 0.712(7)$^{\star}$ & {\text{N/A}} & 1.2\\
\hline
\hline
\end{tabular}
\end{center}
\caption{The numerical values of $\nu$ and $\omega$ calculated from 
$\rho_{s}L$ and $Q_2$ for the herringbone-dimer model on the square lattice 
($(J'/J)_c$ is fixed to 2.4980). 
All results are obtained by using a first order Taylor expansion in $tL^{1/\nu}$ of 
Eq.~(\ref{approx}) with $\omega$ and $b$ left as fitting parameters
except those with a star which are determined through fits with a fixed $\omega = 0.78$.}
\end{table}

\subsection{Results for the herringbone-dimer spin-1/2 Heisenberg model on 
the square lattice}
After having calculated $(J'/J)_c$ and $\nu$ 
for the phase transition induced by dimerization of the staggered-dimer 
spin-1/2 Heisenberg model on the honeycomb lattice, we turn to
investigating the corresponding critical theory of the herringbone-dimer model on the square lattice. Since for this model one has
$\rho_{s1} = \rho_{s2}$, the relevant observables used
in our finite-size scaling analysis are $\rho_{s}L$, which is the average of $\rho_{s1}L$
and $\rho_{s2}L$, and the second Binder ratio $Q_2$ (figure 3). 
To calculate $\nu$, we first carry out several analysis
by employing the second order Taylor expansion in $tL^{1/\nu}$ of Eq.~(\ref{approx}), with the subleading
correction included explicitly, to fit our Monte Carlo data of $\rho_{s}L$ with variant range of $L$.
Remarkably, a numerical value of $\nu$ compatible with $\nu=0.7112(5)$ can be obtained  
if the smallest and largest box sizes used in the fits are larger than $24$ and $96$, 
respectively. The results of $(J'/J)_c$ and $\nu$ calculated from these fits are 
listed as the first 5 rows in table 2. Further, the values of $\omega$ determined 
from these fits ranges from 0.58 to 0.79 with an average of 0.66. Notice $\omega \sim 0.66$ 
we obtain is slightly bellow the expected $\omega \sim 0.78$ in the $O(3)$ universality 
class, hence is consistent with the scenario suggested in \cite{Fritz11}. In particular,
the correction to scaling due to the cubic term introduced in \cite{Fritz11} 
reflects in the value of $\omega$. This observation is in agreement with what
we have obtained for the staggered-dimer model on the honeycomb lattice in previous section.
However, these results for $\omega$ should only be considered as effective ones.
Similarly, a fit using $Q_2$ with $24 \le L \le 136$ as well as a second order Taylor 
expansion in $tL^{1/\nu}$ of Eq.~(\ref{approx}), with the confluent correction left as fitting
parameters for the fit, leads to $(J'/J)_c = 2.49800(7)$
and $\nu=0.714(4)$. Notice the determined $\nu = 0.714(4)$ is consistent with
$\nu = 0.7112(5)$. Further, the confluent exponent $\omega$ from the fit is given by
$\omega = 2.0(2)$. Finally, while using other range of $L$ for $Q_2$ we can arrive
at values of $\nu$ agreeing with $\nu=0.7112(5)$, the $\omega$ calculated from these
additional fits are poor determined. Notice that a second order Taylor expansion 
in $tL^{1/\nu}$ of Eq.~(\ref{approx}) with the subleading correction included for 
the fit contains seven fitting parameters, 
which is at the border of reasonable amount of the unknown coefficients for a fit. 
Still, one would like to understand whether a consistent $\nu$ with $\nu=0.7112(5)$ 
can be obtained from the fits with fewer fitting parameters. Interestingly, using
$(J'/J)_c = 2.4980$, the data points very close to the critical point, as well as
a first order Taylor expansion in $tL^{1/\nu}$ of Eq.~(\ref{approx}) with the confluent 
correction included explicitly (which has five unknown coefficients), 
the values of $\nu$ determined from these new fits for both $\rho_{s}L$ and $Q_2$ are compatible with $\nu=0.7112(5)$ 
(table 3 and top panel of figure 4). Therefore we conclude that, our data points of $\rho_{s}L$ 
and $Q_2$ for the herringbone-dimer model on the square lattice indeed can be
described nicely with the expected $\nu=0.7112(5)$ in the $O(3)$ universality
class. Notice that the values of $\omega$ calculated from the additional fits 
(first order Taylor expansion of Eq.~(\ref{approx})) related to $\rho_{s}L$
has an average of $0.55$, hence again is in agreement with the scenario of a 
large correction to scaling for this phase transition induced by spatial 
anisotropy.
 
Interestingly, while the results we obtain so far are in consistence with 
the scenario that the cubic irrelevant term, which results in the observed enhanced correction 
to scaling, has impact on the confluent exponent $\omega$,
the numerical values of $\nu$ determined from the fits with a fixed $\omega = 0.78$
are also compatible with $\nu = 0.7112(5)$ for both $\rho_{s}L$ and $Q$ (table 2 and table 3). 
For instance, a fit 
using a fixed $\omega = 0.78$
to the observable $\rho_{s}L$ with $32 \le L \le 96$ leads to $\nu = 0.707(3)$,
which is in nice agreement 
with the expected result of $\nu=0.7112(5)$. 
Further, we are also able to arrive at values of $\nu$ agreeing with $\nu = 0.7112(5)$
using the first order Taylor expansion in $tL^{1/\nu}$ of Eq.~(\ref{approx}) with a fixed $\omega = 0.78$ 
for the fits (table 3). These additional fits contains only four 
unknown coefficients. Hence, both the strategies of fixing $\omega$ to be 0.78 or 
letting it be a fitting parameter lead to results of
$\nu$ consistent with $\nu=0.7112(5)$. Notice in our earlier calculations
using a second order Taylor expansion of the full ansatz Eq.~(\ref{approx}), 
although the mean average of $\omega$, determined from
the fits with both the $b$ and $\omega$ in Eq.~(\ref{approx}) 
left as fitting parameters, is smaller than the expected 0.78 in most of the cases, 
the uncertainties for $\omega$ from these fits are large. Hence, 
for a spread range $\left(a_1,a_2\right)$ of $\omega$ (i.e. $\omega \in \left(a_1,a_2\right)$), 
consistent $\nu$
with $\nu=0.7112(5)$ is obtained from the fits using the chosen $\omega$ in $\left(a_1,a_2\right)$.
Therefore, it might be premature to conclude that the confluent exponent 
$\omega$ for this phase transition is smaller than 0.78 just from 
what we have obtained so far. In addition, since the values of $\omega$ obtained from the fits might be 
contaminated by higher order terms, a more sophisticated determination of
$\omega$ should be performed. Indeed, as we will demonstrate later, by considering
higher order corrections, the value of $\omega$ determined from the $Q_2$ data points of the 
herringbone-dimer model agrees reasonably well with $\omega \sim 0.78$. 

\begin{figure}
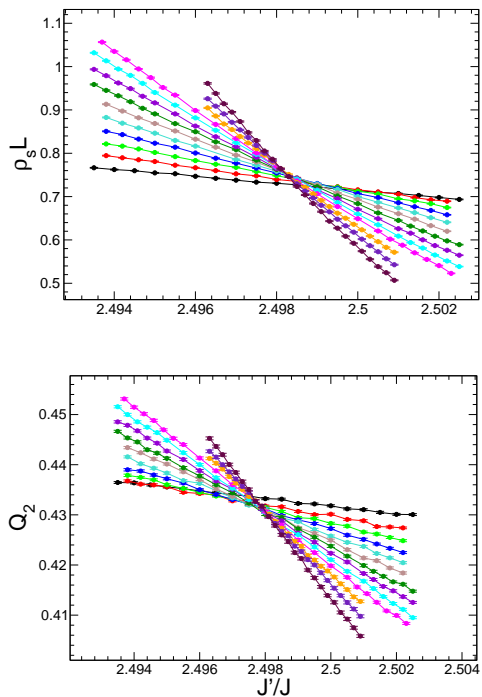

\label{fig3}
\begin{center}
\vbox{
\includegraphics[width=0.35\textwidth]{herringbone_rhosL_1.eps}
\vskip0.47cm
\includegraphics[width=0.35\textwidth]{herringbone_Q2_1.eps}
}
\end{center}
\caption{Monte Carlo data of $\rho_{s}L$ and $Q_2$ with $24 \le L \le 136$ for 
the herringbone-dimer spin-1/2 Heisenberg model on the square lattice. 
}
\end{figure}

\begin{table}
\label{tab3}
\begin{center}
\begin{tabular}{ccccc}
\hline
{\text{observable}}& $\,L\, $  & $\nu$ & $(J'/J)_c$ & $\chi^2/{\text{DOF}}$\\
\hline
\hline
$\rho_{s1}2L$  & $24\, \le \,L\, \le\, 136$  & 0.707(2) & 1.90955(9) & 1.1\\
\hline
$\rho_{s1}2L$  & $24\, \le \,L\, \le\, 96$  & 0.705(3) & 1.9096(2) & 1.2\\
\hline
$\rho_{s1}2L$  & $24\, \le \,L\, \le\, 72$  & 0.696(5) & 1.9095(6) & 1.1\\
\hline
$\rho_{s1}2L$  & $32\, \le \,L\, \le\, 136$  & 0.708(2) & 1.90956(12) & 1.1\\
\hline
$\rho_{s1}2L$  & $32\, \le \,L\, \le\, 96$  & 0.706(3) & 1.9097(2) & 1.1\\
\hline
$\rho_{s1}2L$  & $24\, \le \,L\, \le\, 136$  & 0.706(2)$^{\star}$ & 1.90960(3)${^\star}$ & 1.2\\
\hline
$\rho_{s1}2L$  & $32\, \le \,L\, \le\, 72$  & 0.698(5)$^{\star}$ & 1.90956(16)${^\star}$ & 1.1\\
\hline
$\rho_{s1}2L$  & $32\, \le \,L\, \le\, 96$  & 0.707(3)$^{\star}$ & 1.90961(7)${^\star}$ & 1.2\\
\hline
$\rho_{s1}2L$  & $24\, \le \,L\, \le\, 136$  & 0.704(3) & 1.9095(1) & 1.1\\
\hline
$\rho_{s1}2L$  & $32\, \le \,L\, \le\, 136$  & 0.705(3) & 1.90956(13) & 1.1\\
\hline
\hline
\end{tabular}
\end{center}
\caption{The numerical values of $\nu$ and $(J'/J)_c$ calculated from 
$\rho_{s1}2L$ for the ladder-dimer model on the square lattice. 
While the results presented in the first eight rows are obtained by using a second 
order Taylor expansion in $tL^{1/\nu}$ of Eq.~(\ref{approx}), those listed in the last
two rows are determined with a third order Taylor expansion. 
Further, all results are calculated with the $\omega$ and $b$ in Eq.~(\ref{approx}) 
left as fitting parameters except those with a star which are determined through fits
with a fixed $\omega = 0.78$.}
\end{table}

\begin{figure}
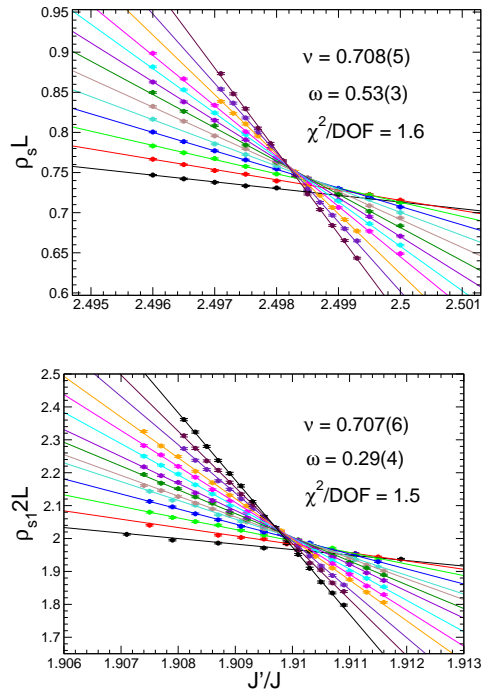

\label{fig4}
\begin{center}
\vbox{
\includegraphics[width=0.35\textwidth]{herringbone_rhosL_near_critical_point.eps}
\vskip0.45cm
\includegraphics[width=0.35\textwidth]{ladder_rhos12L_near_critical_point.eps}
}
\end{center}
\caption{Fits of $\rho_{s}L$ ($24 \le L \le 136$) of the herringbone-dimer model (top panel)
and $\rho_{s1}2L$ ($24 \le L \le 136$)
of the ladder-dimer model (bottom panel) to the first order Taylor expansion in $tL^{1/\nu}$ 
of the full ansatz Eq.~(\ref{approx}).
While the circles are the numerical Monte Carlo data from the simulations, 
the solid curves are obtained by using the results from the fits.}
\end{figure}

\begin{figure}
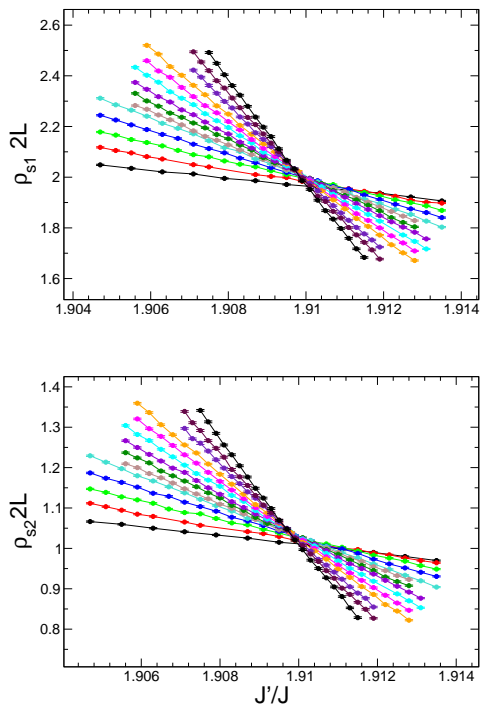

\label{fig4.5}
\begin{center}
\vbox{
\includegraphics[width=0.35\textwidth]{ladder_rhos12L_1.eps}
\vskip0.48cm
\includegraphics[width=0.35\textwidth]{ladder_rhos22L_1.eps}
}
\end{center}
\caption{Monte Carlo data of $\rho_{s1}2L$ and $\rho_{s2}2L$ with $24 \le L \le 136$ for 
the ladder-dimer spin-1/2 Heisenberg model on the square lattice. 
}\vskip0.5cm
\end{figure}

\begin{table}
\label{tab3.5}
\begin{center}
\begin{tabular}{ccccc}
\hline
{\text{observable}}& $\,L\, $  & $\nu$ & $\omega$ & $\chi^2/{\text{DOF}}$\\
\hline
\hline
$\rho_{s1}2L$  & $24\, \le \,L\, \le\, 136$  & 0.707(6) & 0.29(4) & 1.5\\
\hline
$\rho_{s1}2L$  & $32\, \le \,L\, \le\, 136$  & 0.709(6) & 0.26(6) & 1.5\\
\hline
$\rho_{s1}2L$  & $24\, \le \,L\, \le\, 96$  & 0.703(8) & 0.28(6) & 1.5\\
\hline
\hline
\end{tabular}
\end{center}
\caption{The numerical values of $\nu$ and $\omega$ calculated from 
$\rho_{s1}2L$ for the ladder-dimer model on the square lattice 
($(J'/J)_c$ is fixed to 1.9095). 
All results are obtained by using a first order Taylor expansion in $tL^{1/\nu}$ 
of the full ansatz Eq.~(\ref{approx}).}
\end{table}

\subsection{Results for the ladder-dimer spin-1/2 Heisenberg model on 
the square lattice}
The final dimer model considered in this study is the 2-d quantum Heisenberg model
on the square lattice with a ladder spatial anisotropy, which has been studied extensively in 
the literature. Here for completeness, we have re-investigated the phase transition induced 
by dimerization of this model. Intuitively, due to their similarity,
one might expect that the good scaling behavior of the observable $\rho_{s2}2L$,
found for the staggered-dimer model on both the square and honeycomb lattices,
will emerge again for the ladder-dimer model. Interestingly, the effects of
the correction to scaling for $\rho_{s1}2L$ and $\rho_{s2}2L$ of 
this model are about the same qualitatively (figure 4). This indicates the fundamental 
difference regarding the correction to scaling 
between the staggered- and ladder-dimer models as suggested in \cite{Fritz11}.
Similar to the analysis performed for obtaining $\nu$ and $(J'/J)_c$ for 
the herringbone-dimer model, a second order Taylor expansion in $tL^{1/\nu}$ 
of Eq.~(\ref{approx}), with the subleading correction included explicitly, 
is employed to fit the Monte Carlo data of $\rho_{s1}2L$ 
with variant range of $L$. The obtained $(J'/J)_c$ and $\nu$ are in table 4.
Interestingly, table 4 implies that a result for $\nu$ compatible with $\nu=0.7112(5)$ can be obtained  
as well if the smallest and largest box sizes used in the fits are larger than $24$ and $96$, 
respectively. Further, although with considerable large uncertainties, most the values of $\omega$ determined 
from these fits are smaller than $0.78$ and are compatible in magnitude
with those determined from the herringbone-dimer model. 
In addition, using $(J'/J)_c = 1.9095$, the data points very close to the critical point, as well as
a first order Taylor expansion in $tL^{1/\nu}$ of the full ansatz Eq.~(\ref{approx}), 
the values of $\nu$ determined from these new 
fits for $\rho_{s1}2L$ are compatible with $\nu=0.7112(5)$ 
(table 5 and bottom panel of figure 4). Therefore we conclude that, our data points of $\rho_{s1}2L$ 
for the ladder-dimer model on the square lattice indeed 
can be described nicely with the expected $\nu=0.7112(5)$ in the $O(3)$ universality
class. Finally, similar to the results for
the herringbone-dimer model, with a fixed $\omega = 0.78$,
the values of $\nu$ calculated from our finite-size scaling analysis are compatible with
$\nu=0.7112(5)$ as well (table 4).

\section{Determination of the exponent $\beta/\nu$}

After having calculated the critical exponent $\nu$ from the relevant 
observables for the models described by figure 1, we turn to the 
determination of the exponent $\beta/\nu$. 
To calculate $\beta/\nu$, the scaling behavior of
the observables $\langle | m_s^z | \rangle$ and $\langle (m_s^z)^2 \rangle$ 
are studied. Specifically, at critical points and for large $L$, 
the observable $\langle |m_s^z|^k \rangle$  
should scale as
\begin{equation}
\label{beta_nu}
\langle |m_s^z|^k \rangle = (a_k + b_kL^{-\omega})L^{-k\beta/\nu}, 
\end{equation}
where $a_k$, $b_k$ are some constants for $k = 1$ and each even positive integer $k$. 
Since precise knowledge of the critical points is essential in determining the
exponent $\beta/\nu$, we use the values of $(J/J')_c$ obtained in 
previous sections when calculating the critical exponent $\nu$.
Interestingly, as shown in tables 1, 2, and 4, small statistical 
deviation between some of the determined critical points of the same model
is found. We attribute such small discrepancy to the presence of higher
order subleading corrections which are not taken into account in
our analysis, as well as the fact that the bootstrap 
resampling method used in calculating $(J/J')_c$ and $\nu$ might occasionally leads
to underestimated errors. While small deviation is observed, the accuracy of 
$(J/J')_c$ presented in tables 1, 2, and 4, is sufficient for determining $\beta/\nu$ 
by investigating the scaling behavior of
$\langle | m_s^z | \rangle$ and $\langle (m_s^z)^2 \rangle$, at the 
corresponding critical points. Hence, the values of $(J/J')_c$
for the herringbone- and ladder-dimer models 
on the square lattice, as well as the staggered-dimer model on the honeycomb lattice 
are taken to be 2.4980, 1.9095, and 1.7355, respectively. Further, we have carried out
additional simulations at these critical points so that the largest lattice size we reach
for both the herringbone- and staggered-dimer models is $L = 184$. First of all, let us 
focus on the results of $\beta/\nu$ obtained from $\langle | m_s^z | \rangle$. 
Interestingly,
with the expected leading scaling behavior, only from the ladder-dimer 
model we are able to reach a value of $\beta/\nu$ which is in agreement
with the established result $\beta/\nu = 0.519(1)$ in the
literature. For example, while a fit using the leading scaling expectation
and the observable $\langle | m_s^z | \rangle$ with $ L \ge 72$ of 
the ladder-dimer model results in $\beta/\nu = 0.517(2)$ 
(top panel of figure 6), 
the corresponding numerical value of 
$\beta/\nu$ determined from the same observable, with a similar range of $L$, 
is given by $\beta/\nu = 0.527(3)$ ($\beta/\nu = 0.531(3)$) for the 
herringbone-dimer model (staggered-dimer model on the honeycomb lattice). 
Further, using $L \ge 128$ ($L \ge 120$), the value of $\beta/\nu$ obtained from
a fit without the confluent correction for the herringbone-dimer model
(staggered-dimer model on the honeycomb lattice) is given by 
$\beta/\nu = 0.522(5)$ ($\beta/\nu = 0.526(3)$).
From this outcome and in conjunction with the results reached from previous sections, one  
concludes that the correction to scaling for the staggered- and herringbone-dimer models
are indeed enhanced as proposed in \cite{Fritz11}. In particular, 
the effect of the correction to scaling due to the cubic term is the reduction
of the magnitude of the confluent exponent $\omega$.
Surprisingly, for both the herringbone-dimer model on the square lattice 
and the staggered-dimer model on the honeycomb lattice, 
using the data of $\langle | m_s^z | \rangle$ with 
$16 \le L \le 184$, a fit including
the subleading correction and a fixed $\omega = 0.78$ leads to
values of $\beta/\nu$ compatible with $\beta/\nu=0.519(1)$ 
(middle and bottom panels of figure 6). 
Further, if $\omega$ is left as a fitting parameter, 
although consistent $\beta/\nu$ with $\beta/\nu = 0.519(1)$ are obtained
from the fits associated with the herringbone-dimer model, the 
uncertainties for $\beta/\nu$ and $\omega$ are increased 
significantly. For example, the $\beta/\nu$ and $\omega$ determined from a fit using
$\langle | m_s^z | \rangle$ with $16 \le L \le 184$ of the herringbone-dimer model 
are given by $\beta/\nu = 0.521(9)$ and $0.86(50)$, respectively.
Finally, a fit to the observable $\langle | m_s^z | \rangle$ of the ladder-dimer model 
with a fixed $\omega = 0.78$ leads to $ \beta/\nu = 0.516(3)$, which is consistent 
with $\beta/\nu=0.519(1)$ as well. Interestingly, a value of $\beta/\nu$ 
slightly below $\beta/\nu=0.519(1)$ is reached when the leading scaling prediction
is employed to fit all available data of $\langle |m_s^z| \rangle$ of the 
ladder-dimer model. This in turn implies that the coefficient 
$b_1$ in Eq.~(\ref{beta_nu}) for the ladder-dimer model is small in magnitude.
Indeed, the magnitude of $b_1$ obtained from applying the full ansatz 
Eq.~(\ref{beta_nu}) with a fixed $\omega = 0.78$ to $\langle |m_s^z| \rangle$ of the ladder-dimer model
is of order $10^{-2}$ (The uncertainty for $b_1$ is comparable to $b_1$ 
in magnitude as well). Finally, we have also carried out an additional analysis 
with different fixed values of $\omega$ in the fits. The obtained $\beta/\nu$ for 
these additional fits are shown in table 7. From table 7 one concludes that
the values of $\omega$ that would lead to consistent $\beta/\nu$ with 
$\beta/\nu = 0.519(1)$ for both the herringbone- and staggered-dimer models 
ranges from 0.7 to 0.9, which matches reasonable well with the expected
value 0.78. All the results we have reached so far imply that, 
our data points of $\langle | m_s^z | \rangle$ 
for all the three dimerized models depicted in figure 1 are compatible with 
the established value of $\omega \sim 0.78$ in the $O(3)$ universality class.

After having demonstrated that our Monte Carlo data of $\langle | m_s^z | \rangle$, for 
all the three dimerized models investigated here, are compatible with the established results
of $\beta/\nu = 0.519(1)$ and $\omega = 0.78$ in the $O(3)$ universality class,
a similar scenario is reached when considering the observable $\langle ( m_s^z )^2 \rangle$.
For example, using the leading scaling prediction, only from the ladder-dimer model we
can reach a value for $\beta/\nu$ compatible with $\beta/\nu=0.519(1)$ (top panel
of figure 7). Further, with a fixed $\omega = 0.78$, the 
fits for the herringbone-dimer model on the square lattice and the staggered-dimer model on the
honeycomb lattice result in $\beta/\nu = 0.5202(15)$ and $\beta/\nu = 0.518(2)$, respectively 
(middle and bottom panels of figure 7). The results of $\beta/\nu = 0.5202(15)$ and 
$\beta/\nu = 0.518(2)$ we just obtain are in quantitative agreement with the expected
$\beta/\nu=0.519(1)$. Further, an analysis for $\langle ( m_s^z )^2 \rangle$, 
with variant fixed values of $\omega$ in the fits, leads to a similar conclusion like 
that of $\langle | m_s^z | \rangle$. Specifically, from $\langle ( m_s^z )^2 \rangle$, 
the values of $\omega$ that would lead to consistent $\beta/\nu$ with 
$\beta/\nu = 0.519(1)$ for both the herringbone- and staggered-dimer models 
ranges from 0.7 to 0.9 as well (table 8). Tables 6, 7, and 8 summarizes our calculations 
on determining $\beta/\nu$ for the phase transitions induced by dimerization for 
all the three models shown in figure 1.

\begin{figure}
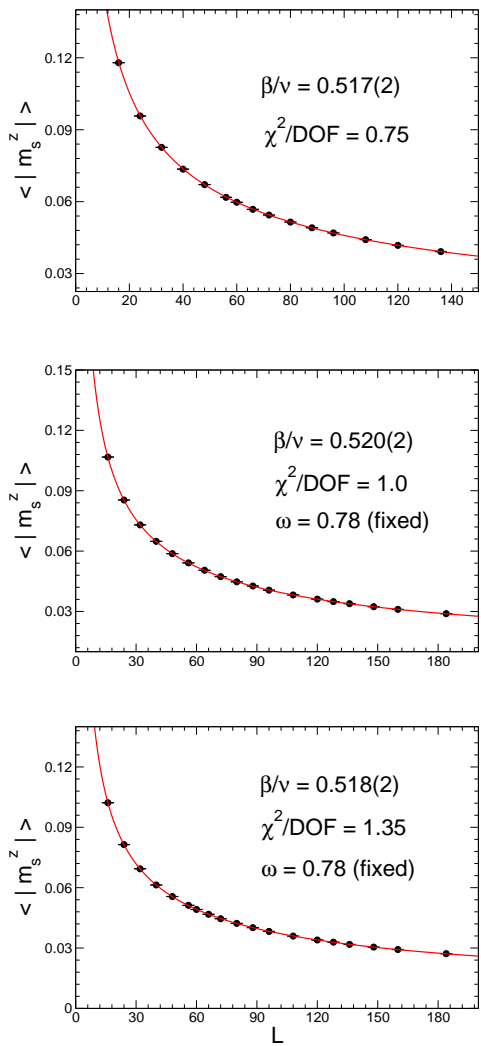

\label{fig5}
\begin{center}
\vbox{
\includegraphics[width=0.35\textwidth]{ladder_stag_mag_beta_nu.eps}
\vskip0.45cm
\includegraphics[width=0.35\textwidth]{herringbone_stag_mag_beta_nu.eps}
\vskip0.46cm
\includegraphics[width=0.35\textwidth]{honeycomb_stag_mag_beta_nu.eps}
}
\end{center}
\caption{Determination of $\beta/\nu$ for the ladder- and herringbone-dimer models (top and middle panels) 
on the square lattice as well as for the staggered-dimer model on the honeycomb lattice (bottom panel). 
While $\beta/\nu=0.517(2)$ for the ladder-dimer model is obtained by fitting the 7
largest $L$ data points of $\langle | m_s^z | \rangle$ to their expected leading scaling behavior, 
fits with a fixed $\omega = 0.78$ to $\langle | m_s^z | \rangle$ of the herringbone-dimer model
on the square lattice and the staggered-dimer model on the honeycomb lattice
result in $\beta/\nu = 0.520(2)$ and $\beta/\nu=0.518(2)$, respectively.}
\end{figure}

\begin{figure}
\label{fig6}
\begin{center}
\vbox{
\includegraphics[width=0.35\textwidth]{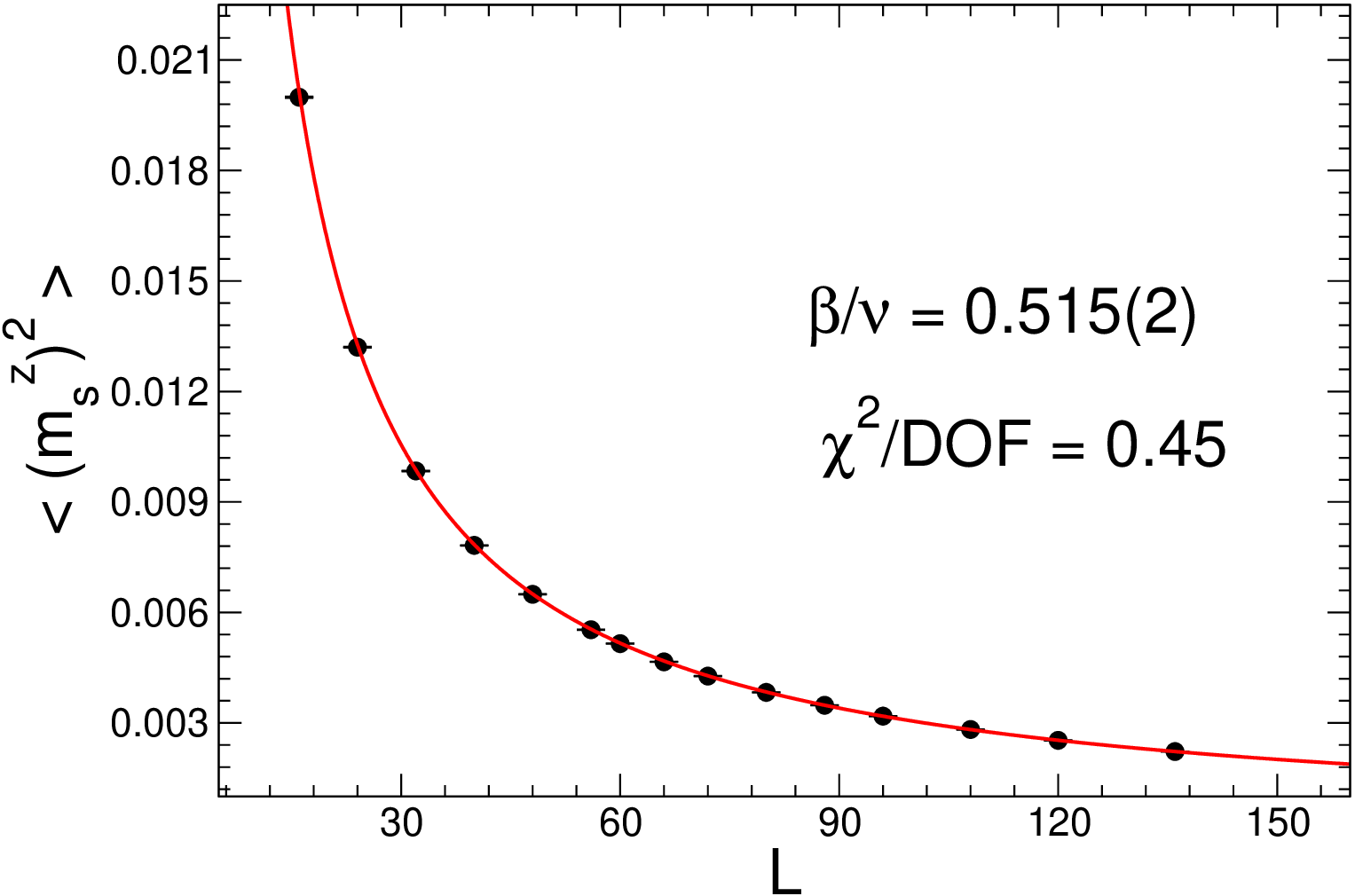}
\vskip0.45cm
\includegraphics[width=0.35\textwidth]{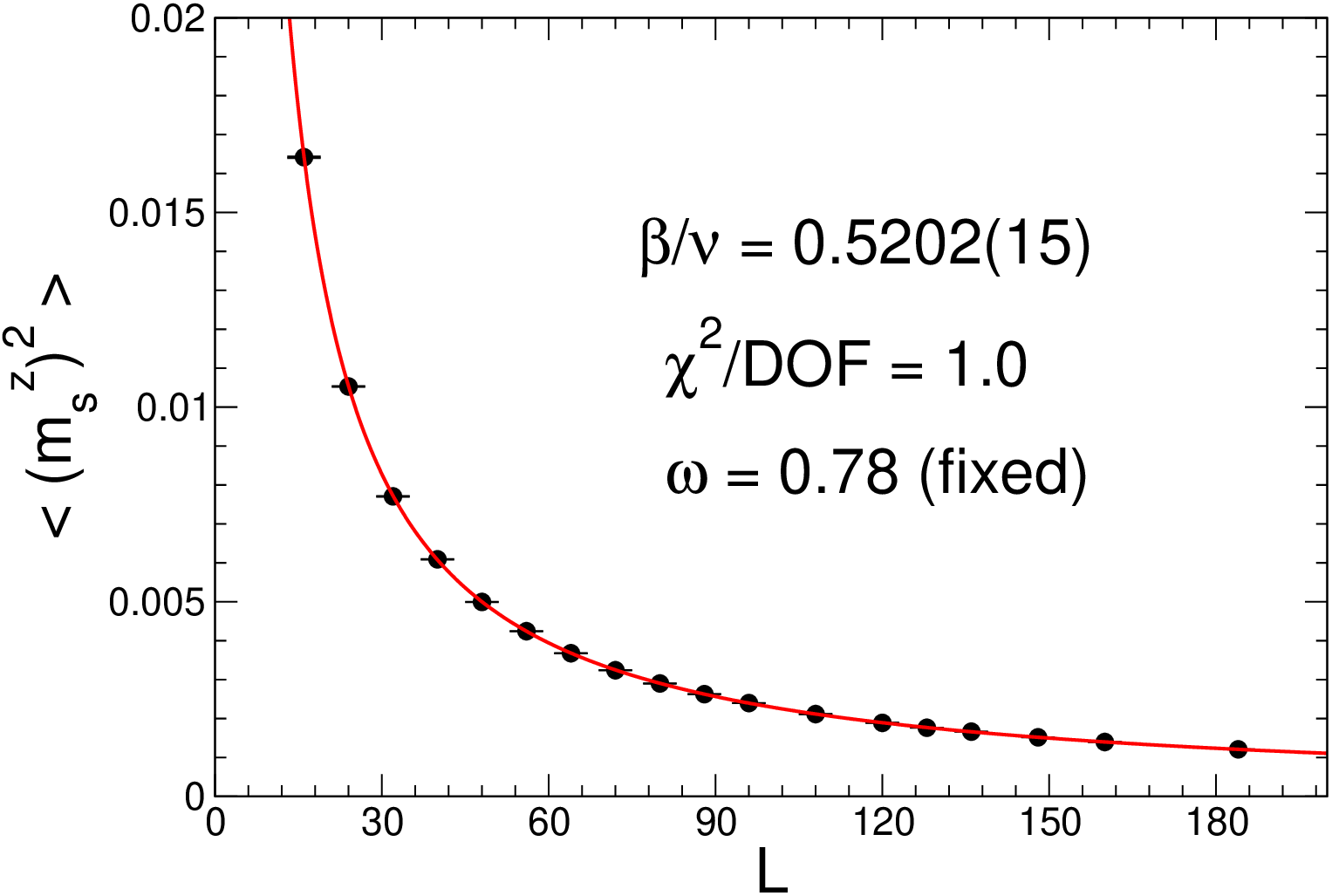}
\vskip0.46cm
\includegraphics[width=0.35\textwidth]{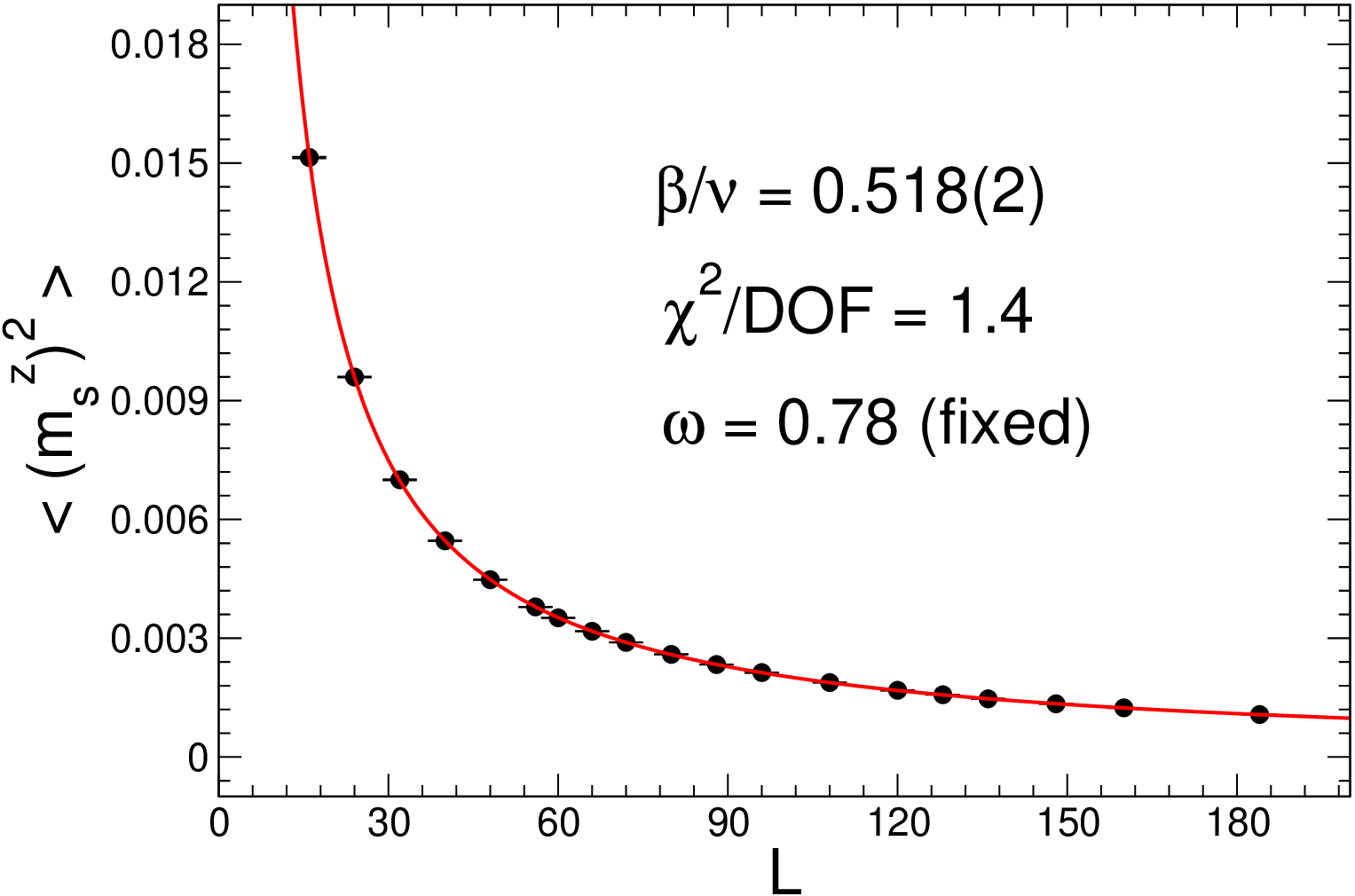}
}
\end{center}
\caption{Determination of $\beta/\nu$ for the ladder- and herringbone-dimer models (top and middle panels) 
on the square lattice as well as for the staggered-dimer model on the honeycomb lattice (bottom panel). 
While $\beta/\nu=0.515(2)$ for the ladder-dimer model is obtained by fitting the 5
largest $L$ data points of $\langle ( m_s^z )^2 \rangle$ to their expected leading scaling behavior, 
fits with a fixed $\omega = 0.78$ to $\langle ( m_s^z )^2 \rangle$ of the herringbone-dimer model
on the square lattice and the staggered-dimer model on the honeycomb lattice
result in $\beta/\nu = 0.5202(15)$ and $\beta/\nu=0.518(2)$, respectively.}
\end{figure}

\begin{table}
\label{tab4}
\begin{center}
\begin{tabular}{ccccc}
\hline
{\text{model}} & $L$ & $\beta/\nu$ & $\chi^2/{\text{DOF}}$\\
\hline
\hline
{\text{ladder}} & $ 72 \le L \le 136$ & 0.517(2)${^\star}$  & 0.75\\
\hline
{\text{ladder}} & $16 \le L \le 136 $ & 0.516(3)  & 1.1 \\
\hline
{\text{herringbone}} & $72 \le L \le 136$ & 0.527(3)${^\star}$ & 1.5\\
\hline
{\text{herringbone}} & $128 \le L \le 184$ & 0.522(5)${^\star}$ & 0.7\\
\hline
{\text{herringbone}}  & $16 \le L \le 184$ & 0.520(2)  & 1.0 \\
\hline
{\text{staggered}}  &  $72 \le L \le 136$ & 0.531(3)$^{\star}$  & 1.2 \\
\hline
{\text{staggered}}  &  $120 \le L \le 184$ & 0.526(3)$^{\star}$  & 0.55 \\
\hline
{\text{staggered}}  & $16 \le L \le 184$  & 0.518(2)  & 1.35 \\
\hline
{\text{ladder}} & $ 88 \le L \le 136$ & 0.515(2)${^\star}$  & 0.45\\
\hline
{\text{ladder}} & $16 \le L \le 136 $ & 0.515(2)  & 1.3 \\
\hline
{\text{herringbone}} & $128 \le L \le 184$ & 0.525(4)${^\star}$ & 1.25\\
\hline
{\text{herringbone}}  & $16 \le L \le 184$ & 0.5202(15)  & 1.0 \\
\hline
{\text{staggered}}  &  $128 \le L \le 184$ & 0.527(4)$^{\star}$  & 1.1 \\
\hline
{\text{staggered}}  & $16 \le L \le 184$  & 0.518(2)  & 1.4 \\
\hline
\hline
\end{tabular}
\end{center}
\caption{The numerical values of $\beta/\nu$ calculated from 
$\langle | m_s^z | \rangle$ (the first 8 rows) and
$\langle ( m_s^z )^2 \rangle$ (the last 6 rows) for the dimerized models 
considered in this study. 
All results are obtained with a fixed $\omega= 0.78$ except those with a star
which are determined by using the expected leading scaling prediction.}
\end{table}

\begin{table}
\label{tab5}
\begin{center}
\begin{tabular}{ccccc}
\hline
{\text{model}} & $\omega$(fixed) & $\beta/\nu$ & $\chi^2/{\text{DOF}}$\\
\hline
\hline
{\text{herringbone}} & 1.0 & 0.523(2) & 1.0\\
\hline
{\text{herringbone}} & 0.9 & 0.522(2) & 1.0\\
\hline
{\text{herringbone}}  & 0.7 & 0.5181(23)  & 1.06 \\
\hline
{\text{herringbone}}  &  0.65 & 0.5168(25)  & 1.05 \\
\hline
{\text{herringbone}}  &  0.6 & 0.5152(27)  & 1.04 \\
\hline
{\text{herringbone}}  & 0.45  & 0.5078(36)  & 1.05 \\
\hline
{\text{staggered}} & 1.0 & 0.523(2)  & 1.55\\
\hline
{\text{staggered}} & 0.9 & 0.521(2)  & 1.46 \\
\hline
{\text{staggered}} & 0.7 & 0.5150(25) & 1.3\\
\hline
{\text{staggered}}  & 0.65 & 0.5127(25)  & 1.23 \\
\hline
{\text{staggered}}  &  0.6 & 0.5101(25)  & 1.2 \\
\hline
{\text{staggered}}  & 0.45  & 0.4974(35)  & 1.15 \\
\hline
\hline
\end{tabular}
\end{center}
\caption{The numerical values of $\beta/\nu$ calculated from 
$\langle | m_s^z | \rangle$ ($16 \le L \le 184$) with variant fixed $\omega$ for the fits.}
\end{table}

\begin{table}
\label{tab6}
\begin{center}
\begin{tabular}{ccccc}
\hline
{\text{model}} & $\omega$(fixed) & $\beta/\nu$ & $\chi^2/{\text{DOF}}$\\
\hline
\hline
{\text{herringbone}} & 1.0 & 0.5232(12) & 1.03\\
\hline
{\text{herringbone}} & 0.9 & 0.5221(13) & 1.0\\
\hline
{\text{herringbone}}  & 0.7 & 0.5185(15)  & 1.0 \\
\hline
{\text{herringbone}}  &  0.65 & 0.5172(16)  & 1.0 \\
\hline
{\text{herringbone}}  &  0.6 & 0.5156(18)  & 1.05 \\
\hline
{\text{herringbone}}  & 0.45  & 0.5082(25)  & 1.1 \\
\hline
{\text{staggered}} & 1.0 & 0.5235(15)  & 1.7\\
\hline
{\text{staggered}} & 0.9 & 0.5213(15)  & 1.55 \\
\hline
{\text{staggered}} & 0.7 & 0.515(2) & 1.4\\
\hline
{\text{staggered}}  & 0.65 & 0.512(2)  & 1.35 \\
\hline
{\text{staggered}}  &  0.6 & 0.509(2)  & 1.35 \\
\hline
{\text{staggered}}  & 0.45  & 0.4920(36)  & 1.5 \\
\hline
\hline
\end{tabular}
\end{center}
\caption{The numerical values of $\beta/\nu$ calculated from 
$\langle ( m_s^z )^2 \rangle$ ($16 \le L \le 184$) with variant fixed $\omega$ for the fits.}
\end{table}

\begin{figure}
\label{fig7}
\begin{center}
\includegraphics[width=0.35\textwidth]{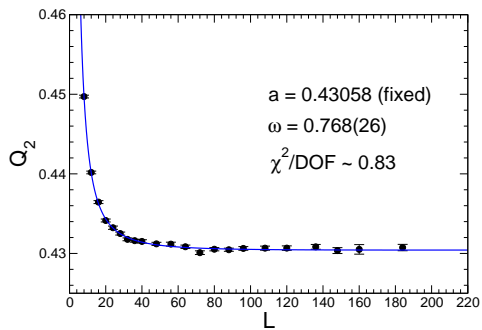}
\end{center}
\caption{Fit of $Q_2$ data ($8 \le L \le 184$) of the herringbone-dimer model 
to Eq.~(7).}
\end{figure}

\section{Determination of the exponent $\omega$}
In previous sections, the numerical values of $\omega$ 
are poor determined. Here we attempt to have better estimates for the confluent exponent $\omega$
of both the herringbone- and staggered-dimer models. 
As a first step toward fulfiling this purpose, let us reanalyze our $\langle |m_s^z| \rangle$ data of 
the herringbone- and staggered-dimer
models from another point of view. Notice the values of $\omega$ and $\beta/\nu$ are poor determined
when both of them are included as fitting parameters for the fits. 
Interestingly, for both the herringbone- and staggered-dimer models, if we fix $\beta/\nu$ to 
the known result $\beta/\nu = 0.519$ in the literature, then the $\omega$ we obtain from these new fits
are much better determined and are in nice agreement with $\omega \sim 0.78$. 
For example, for the herringbone-dimer (staggered-dimer)
model, using a fixed $\beta/\nu = 0.519$, 
we arrive at $\omega = 0.74(8)$ ($\omega = 0.80(5)$). 
Summarizing all the analysis done so far related to the observables $\langle |m_s^z| \rangle$ and 
$\langle (m_s^z)^2 \rangle$, we conclude that our these 
data points are fully compatible with the known results of $\beta/\nu = 0.519(1)$ 
and $\omega \sim 0.78$ in the $O(3)$ universality class.
Next, we will focus on the finite-size scaling analysis of $Q_2$ and $\rho_{s}L$ 
at the corresponding critical points. First of
all, let us discuss the results for the herringbone-dimer model. Notice, at critical points, 
the finite-size scaling ansatz for $Q_2$ (and $\rho_{s}L$ as well)
is given as
\begin{equation}
Q_2 = a + bL^{-\omega} + cL^{-2\omega} + ...
,\end{equation} 
where $a,b,c$ are some constants and ``$...$'' in Eq.~(7) stands for higher order corrections. 
In additional to the corrections associated with the confluent exponent $\omega$, 
there are other subleading corrections with exponents $\omega' > \omega$. Notice that
since the established values of $\omega'$ are larger or equal to $2\omega$, it is
reasonable to employ the scaling ansatz Eq.~(7) for data analysis.  
From our data of $Q_2$ for the herringbone-dimer model, 
we observe two interesting results as follows. First, $Q_2$ converges rapidly to
a constant. Indeed, for $64 \le L \le 184$, an average of the corresponding values of $Q_2$ leads
to $Q_2 = 0.43058(22)$, where the quoted error is obtained by the variance of these $Q_2$.
Second, we find that when the ansatz with only the leading correction is employed for the fits, 
then the values of $\omega$ calculated are much larger than the expected $\omega \sim 0.78$ in the $O(3)$ universality class. For example, a fit using 
Eq.~(7) up to the term $bL^{-\omega}$ and the data points of $4 \le L \le 136$ ($8 \le L \le 120$) leads 
to $\omega \sim 1.74$ ($\omega \sim 1.69$), with a 
$\chi^2/{\text{DOF}} = 1.29$ ($\chi^2/{\text{DOF}} = 1.2$). These results indicate that
the influence of higher order terms already sets in. Hence, instead of using the ansatz with only
the leading correction term, the term of $cL^{-2\omega}$ in Eq.~(7) should be additionaly included in the fits. 
Remarkably, with a fixed $a = 0.43058$ in Eq.~(7), the $\omega$ determined from the fits employing 
the ansatz $0.43058 + bL^{-\omega} + cL^{-2\omega}$ are compatible with the expected 
$\omega \sim 0.78$. For instance, from a fit using the $Q_2$ data with $8 \le L \le 184$, we arrive at
$\omega = 0.768(26)$ which agrees nicely with $\omega = 0.78(1)$. The top 6 rows of table 9 summarize the results of 
these fits employing the $Q_2$ data with different range of $L$, and fig.~8 demonstrates one outcome of these fits. Notice in table 9, 
if the coefficient $a$ in Eq.~(7) is left as a fitting parameter, then from the $Q_2$ data, the values of 
$\omega$ are less accurately determined and are with large errors, but still they are in reasonable agreement with $\omega \sim 0.78$.
Similarly, by considering the observable $\rho_{s}L$ at the critical point, 
fits of applying Eq.~(7) to $\rho_{s}L$ data points lead to consistent $\omega$ with 
$\omega = 0.78(1)$, see the bottom 4 rows of table 9 as well as fig.~9. 
Interestingly, when the $a$ in Eq.~(7) is left as an unknown parameter for the fits, 
then the values of $\omega$, obtained from the observable $\rho_{s}L$, are in better agreement with $\omega \sim 0.78$ 
than those determined from $Q_2$. From our finite-size scaling analysis
on $Q_2$ and $\rho_{s}L$ at the critical point, we conclude that in previous section, the obtained values of $\omega$ are deviated from 
$\omega \sim 0.78$ significantly because they are contaminated by higher order corrections. This is confirmed 
by the fact that, our reinvestigation of 
$\rho_{s}L$ ($Q_2$) data close to the critical point, using first order Taylor expansion of Eq.~(5) as well as including the
correction $b_1L^{-2\omega}$ in the fitting ansatz, leads to $\nu= 0.709(3)$, $\omega \sim 0.8$ ($\nu=0.710(5)$, $\omega \sim 0.82$)
with a $\chi^2/{\text{DOF}} = 1.57$ ($\chi^2/{\text{DOF}} = 0.82$). In conclusion, for the herringbone-dimer model, the values
of $\omega$ we obtain from the finite-size scaling analysis are in quantitative consistence with the known $O(3)$ result $\omega = 0.78(1)$.

After having shown that the numerical values of $\omega$ calculated from the $Q_2$ and $\rho_{s}L$ data points of the herringbone-dimer model
agree nicely with the expected $\omega \sim 0.78$ in the $O(3)$ universality class, we turn to determining the exponent $\omega$
of the staggered-dimer model on the honeycomb lattice. The observables $Q_2$ and $\rho_{s}L$ are used as well. Interestingly, while
for the herringbone-dimer model, the observable $Q_2$ converges rapidly to a constant, this is not the case for
the staggered-dimer model. As a result, $a$ in Eq.~(7) must be left as one fitting parameter. 
Table 10 summarizes the results of these fits associated with the staggered-dimer model, and
fig.~10 shows one outcome of these fits. 
Notice the obtained results of $\omega$ in table 10 are slightly below $\omega \sim 0.78$ (This also occurs for some
results associated with the herringbone-dimer model). It is anticipated that
the deviation between the values of $\omega$ in table 10 and $\omega \sim 0.78$ will not lead to
the large correction to scaling known in the literature for this model. Hence 
we attribute the observed small differences between the results of $\omega $ shown in table 10 and $\omega = 0.78(1)$
to higher order corrections not taken into account in our analysis.
Surprisingly, when we perform a similar analysis for the observable $\rho_{s1}L$, the values of $\omega$ we obtain 
are significantly lower than $\omega \sim 0.78$. There are several possible explanations for this 
observation, for instance, $\rho_{s1}L$ is sensible to the critical point.
A thorough determination of $\omega$ for the staggered model,
including considering the uncertainties of the critical point, studying other relevant observables, 
as well as employing the idea of fixing the aspect ratio of winding numbers squared (Notice the spatial winding numbers squared in 1- and 2-directions
take the same values automatically for the herringbone-dimer model), will be left for future work.

\begin{figure}
\label{fig8}
\begin{center}
\includegraphics[width=0.35\textwidth]{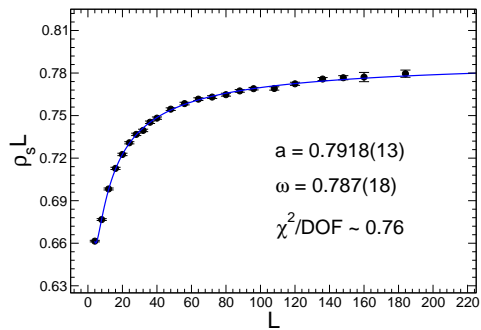}
\end{center}
\caption{Fit of $\rho_{s}L$ data ($4 \le L \le 184$) of the herringbone-dimer model 
to Eq.~(7).
}
\end{figure}

\begin{table}
\label{tab9}
\begin{center}
\begin{tabular}{ccccc}
\hline
Observable & $\,L\, $ & $a$ &  $\omega$ & $\chi^2/{\text{DOF}}$\\
\hline
\hline
$Q_2$&  $8\, \le \,L\, \le\, 184$ & 0.43058(fixed) & 0.768(26) & 0.83\\
\hline
$Q_2$& $8\, \le \,L\, \le\, 120$ & 0.43058(fixed) & 0.760(26) & 0.89\\
\hline
$Q_2$& $8\, \le \,L\, \le\, 96$ & 0.43058(fixed)& 0.751(26)  & 0.92\\
\hline
$Q_2$& $8\, \le \,L\, \le\, 184$ & 0.4312(6) & 0.693(56) & 0.78\\
\hline
$Q_2$& $8\, \le \,L\, \le\, 120$ & 0.4310(7) & 0.708(67) & 0.92\\
\hline
$Q_2$& $8\, \le \,L\, \le\, 96$ & 0.4307(7)&  0.738(82) & 1.0\\
\hline
$\rho_s L$& $4\, \le \,L\, \le\, 184$ &0.7918(13) & 0.787(18) & 0.76\\
\hline
$\rho_s L$& $4\, \le \,L\, \le\, 136$ &0.7912(14) & 0.794(19) & 0.81\\
\hline
$\rho_s L$& $8\, \le \,L\, \le\, 136$ & 0.794(3)& 0.732(57)  & 0.76\\
\hline
$\rho_s L$& $8\, \le \,L\, \le\, 88$ & 0.794(5)& 0.737(77)  & 0.66\\
\hline
\hline
\end{tabular}
\end{center}
\caption{The numerical values of $\omega$, calculated from 
the observables $Q_2$ and $\rho_{s}L$ at the critical point, for the herringbone-dimer model on the square lattice.
The ansatz used for the fits is $a+bL^{-\omega}+cL^{-2\omega}$.}
\end{table}

\begin{table}
\label{tab10}
\begin{center}
\begin{tabular}{ccccc}
\hline
Observable & $\,L\, $ & $a$ &  $\omega$ & $\chi^2/{\text{DOF}}$\\
\hline
\hline
$Q_2$&  $4\, \le \,L\, \le\, 136$ & 0.4317(3) & 0.719(12) & 0.98\\
\hline
$Q_2$& $4\, \le \,L\, \le\, 88$ & 0.4315(4) & 0.725(13) & 1.04\\
\hline
$Q_2$& $4\, \le \,L\, \le\, 184$ & 0.4319(3)& 0.714(10)  & 0.98\\
\hline
$Q_2$& $6\, \le \,L\, \le\, 136$ & 0.4321(5) & 0.698(21) & 0.96\\
\hline
$Q_2$& $6\, \le \,L\, \le\, 88$ & 0.4319(6) & 0.704(25) & 1.05\\
\hline
$Q_2$& $6\, \le \,L\, \le\, 184$ & 0.4324(5) & 0.688(18) & 0.92\\
\hline
\hline
\end{tabular}
\end{center}
\caption{The numerical values of $\omega$, calculated from 
the observable $Q_2$ at the critical point, 
for the staggered-dimer model on the honeycomb lattice. The ansatz used for the fits
is $a+bL^{-\omega}+cL^{-2\omega}$.}
\end{table}

\begin{figure}
\label{fig9}
\begin{center}
\includegraphics[width=0.35\textwidth]{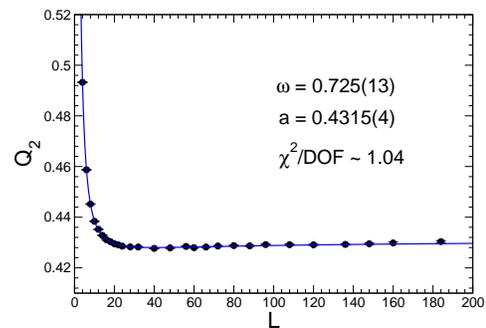}
\end{center}
\caption{Fit of $Q_2$ data of the staggered-dimer model 
to Eq.~(7). The results is obtained by using the data with $4 \le L \le 88$.
}
\end{figure}

\section{Discussion and Conclusion}
In this study, we investigate the phase transitions induced by dimerization for the herringbone- and ladder-dimer
spin-1/2 Heisenberg model on the square lattice, as well as the staggered-dimer model on the honeycomb lattice. 
In particular, we determine the values of the exponents $\nu$ and $\beta/\nu$ with 
high accuracy by employing the finite-size scaling analysis to the relevant observables.
Similar to the scenario found for the staggered-dimer model on the square lattice, 
while the observable $\rho_{s1}2L$ of the staggered-dimer model on the 
honeycomb lattice receives a sizable correction to its scaling, 
the observable $\rho_{s2}2L$ shows a good scaling behavior. 
As a result, using the data points of $\rho_{s2}2L$ with moderate lattice sizes 
as well as the corresponding leading finite-size scaling ansatz (letting $b=0$ in Eq.~(\ref{approx})),
we are able to obtain a value for $\nu$ consistent with the expected 
$\nu=0.7112(5)$ in the $O(3)$ universality class.  
To understand this observation for the staggered-dimer model on both the square and honeycomb lattices
from field theory aspect, is an interesting and important topic to explore. In particular, whether
the cubic term introduced in \cite{Fritz11} is responsible for this unexpected result should be
investigated. Further, while it is argued in \cite{Fritz11} that the herringbone-dimer model belongs
to the category of models receiving a large correction, and our investigation 
supports this scenario, we find that the cubic term most likely has little influence on the 
confluent exponent $\omega$. In particular, our Monte Carlo data of  
$\rho_{s}L$ as well as $Q_2$ are compatible with the established 
result of $\omega \sim 0.78$ in the $O(3)$ universality class, namely with a fixed
$\omega = 0.78$, we are able to arrive at consistent $\nu$
with $\nu=0.7112(5)$ from both $\rho_s L$ and $Q_2$ of the herringbone-dimer model. 
In order to clarify whether the cubic term introduced in \cite{Fritz11} has
no impact on the numerical value of $\omega$, it will be desirable to carry out a more detailed 
investigation to determine $\omega$ with high precision. 
In particular, the consistence of the $\nu$, obtained from the fits using a fixed $\omega = 0.78$, with
$\nu=0.7112(5)$ as shown in tables 2, 3, 4 and 5, is unlikely a
coincidence considering the fact that the conclusion is valid for both the observables spin stiffness and second Binder ratio
of both the herringbone- and ladder-dimer models. 
In addition, the finite-size scaling analysis performed
for the determination of $\beta/\nu$ suggests that, our data points for all the 2-d dimerized models 
with spatial anisotropy considered here are compatible with the established results of 
$\beta/\nu = 0.519(1)$ and $\omega \sim 0.78$
in the $O(3)$ universality class as well. Finally, the consistence of the $\beta/\nu$, determined
from the fits using a fixed $\omega \in \{0.7,0.9 \}$, with $\beta/\nu = 0.519(1)$ implies that the observed
enhanced correction to both the staggered- and herringbone-dimer models is because of the nonuniversal
coefficients $b_k$ in Eq.~(\ref{beta_nu}). Indeed, in tables 6, 7, and 8, 
the values of $b_1$ and $b_2$ determined from the fits associated
with the herringbone- and staggered-dimer models are at least several times 
larger in magnitude than those of the ladder-dimer model (This conclusion remains valid when considering the data sets
generated using lower temperatures). 
It is interesting to notice
that the slopes of $\chi_u/T$ as functions of $T/J$ when approaching
the low-temperature regime ($\chi_u$ is the uniform susceptibility), 
as shown in the
figure 6 of \cite{Fritz11}, imply that the correction for the
staggered- and herringbone-dimer models are large when compared to those
of the ladder- and bilayer-dimer models. This can be considered as a analogy to
our results for $\langle |m_s^z| \rangle$ and $\langle (m_s^z)^2 \rangle$,
and might be used as another evidence for the scenario that the influence on
the scaling due to the cubic term is the amplification of the nonuniversal
prefactors appeared in the scaling forms.
Whether there is a subtlety behind the results for $\langle | m_s^z | \rangle$ and 
$\langle (m_s^z)^2 \rangle$ shown here and the uniform susceptibilities presented in \cite{Fritz11},
or it is just a coincidence should be investigated analytically. 
Indeed, we have shown that the values of $\omega$ for the 
herringbone- and staggered-dimer model agree reasonably well with $\omega \sim 0.78$,
by studying the scaling behavior of $Q_2$ and $\rho_s L$ at the corresponding critical points. 
The rapid saturation of the observable $Q_2$ to a constant is crucial in leading to a precise determination
of $\omega$ for the herringbone-dimer model. While the accuracy for $\omega$ presented here
has not reached the same level of $\omega = 0.78(1)$, we have obtained sufficiently good precision
for $\omega$ to draw the above conclusions. 
Our study of calculating the values of $\omega$ through $Q_2$
and $\rho_{s}L$, at least for the herringbone-dimer model,
at the critical points reinforces our proposed scenario that the enhanced correction
to scaling manifests itself as amplification of the nonuniversal prefactors appeared in the scaling forms.  
While we demonstrate strong evidence that for the herringbone- and staggered-dimer models, 
the exponents $\nu$, $\beta/\nu$, as well as $\omega$ agree quantitatively with the established 
results in the $O(3)$ universality class, still, our estimates for the numerical value of $\omega$
are of a few percent uncertainties. Further, the exponent $\omega$ is associated with the
correction to scaling, and to accurately determine its value is of highly nontrivial.  
Hence one cannot rule out the scenario
that indeed the values of $\omega$ is reduced (slightly) due to the cubic term. 
Hence it is desirable to obtain high statistics data points in order to reach a even higher
precision determination of $\omega$ for the models investigated here \cite{Has99}. 
In light of this, as well as the fact that the values of 
$\omega$ calculated from $\rho_{s1}L$ of the staggered-dimer model are lower than (and statistically different from) 
the expected $\omega \sim 0.78$, a more detailed numerical study (to determine the confluent exponent $\omega$)
and a better theoretical understanding, for the critical theories of the phase transitions investigated here will 
be very useful.

\section{Acknowledgments}
We thank S.~Wessel for useful correspondence.
Partial support from NSC (Grant No. NSC 99-2112-M003-015-MY3)
and NCTS (North) of R.O.C. is acknowledged.

\end{document}